# Frequency Conversion of Entangled State

Aihong Tan, Xiaojun Jia, Changde Xie*

State Key Laboratory of Quantum Optics and Quantum Optics Devices, Institute of Opto-Electronics, Shanxi University Taiyuan 030006, Peoples's Republic of China

The quantum characteristics of sum-frequency process in an optical cavity with an input signal optical beam, which is a half of entangled optical beams, are analyzed. The calculated results show that the quantum properties of the signal beam can be maintained after its frequency is conversed during the intracavity nonlinear optical interaction. The frequency-conversed output signal beam is still in an entangled state with the retained other half of initial entangled beams. The resultant quantum correlation spectra and the parametric dependences of the correlations on the initial squeezing factor, the optical losses and the pump power of the sum-frequency cavity are calculated. The proposed system for the frequency conversion of entangled state can be used in quantum communication network and the calculated results can provide direct references for the design of experimental systems.

42.50.Dv, 03.67.Mn, 03.65.Ud

## 1 Introduction

Quantum entanglement of amplitude and phase quadratures of optical fields, a typical continuous variable(CV) entanglement, has been extensively applied in the quantum information and communication[1]. The unconditionalness of CV entanglement, which is usually generated from the nonlinear optical interaction of a laser with a crystal in a determinant fashion for a given experimental system, is a valuable feature for efficiently exploiting the entanglement resource. The successful experiments of unconditional quantum teleportation, quantum dense coding, quantum entanglement swapping and elementary quantum communication networks based on CV entanglement[2,3,4,5,6,7,8] enhance the interest to explore the schemes establishing more complicated CV quantum communication network and developing CV quantum telecommunication. Recently, a direct quantum interface for photonic qubits at different wavelengths was experimentally demonstrated[9]. In their experiment, the energy-time entanglement of a photon at 1,310 nm wavelength with a photon at 1,550 nm, was coherently transferred to another photon at a wavelength of 710 nm via a process of sum-frequency generation (SFG). Since 710 nm wavelength is close to that of alkaline atomic transitions, the conversed photon can be considered to be used for the storage and processing of quantum information. It is important to preserve the initial entanglement after the wavelength of the light is conversed for building a complete quantum information network of CV using the entangled state of light. In 1990, P.Kumar proposed a scheme for quantum frequency conversion (QFC) and then experimentally proved that the nonclassical intensity correlation can be preserved after the frequency of one of the initial twin beams was conversed[10,11]. In Ref. [10]

and [11], a mode-locked, Q-switched, and frequency-doubled Nd-doped yttrium-aluminum-garnet (Nd:YAG) laser was used as the pump laser for both twin-beam generation and the QFC in a nonlinear crystal.

In the other hand, the stable entangled state of amplitude and phase quadratures of continuous optical fields have been produced through the optical parametric amplification(OPA) processes of a continuous wave (CW) in an optical cavity and applied in a variety of CV quantum information[1-8]. Especially, CW nondegenerate OPA (NOPA) using type II nonlinear $\chi^{(2)}$ crystal can directly provide either the bright entangled optical beams with the correlated amplitude quadratures and the anticorrelated phase quadratures when it operates at amplification[12] or that with the anticorrelated amplitude and the correlated phase at deamplification[6-8]. In this paper, we will discuss the QFC of a CW, which is a half of the Eistain-Podolsky-Rosen (EPR) entangled optical beams produced from a CW NOPA. The optical process used for the QFC is an intracavity SFG. Our calculation shows that the quantum entanglement characteristics of the initial entangled beams can be preserved after the frequency of one of the entangled beams is conversed. The parameter dependences of the preserved quantum entanglement upon the initial squeezing factor, the quality of optical cavity for SFG and the pump power are calculated.

The paper is organized as follows. In the second section, a physical system for the QFC is summarized. Then in the third section, we recall the expressions of EPR entanglement between the amplitude and phase quadratures of the entangled optical beams. The process of SFG is described in the fourth section and the entanglement characteristics between the frequency conversed beam and the retained initial beam are discussed in the fifth section. Finally, we give a brief conclusion in the sixth section.

## 2 Physical system for quantum frequency conversion

The schematic physical system for the QFC is shown in Fig.1. At first, a pair of EPR entangled optical beams with degenerate frequency, $a_1(\omega_1)$ and $a_2(\omega_1)$, are produced from the NOPA via a frequency-down-conversion process of the pump field $a_0(\omega_0)$ at the frequency $\omega_0 = 2\omega_1$ [12]. The NOPA is implemented in an optical cavity involving a type-II phase-matching $\chi^{(2)}$ nonlinear crystal. The $a_{01}(\omega_1)$ and $a_{02}(\omega_1)$ in the coherent state are the injected signals which are polarized in same orientation with $a_1(\omega_1)$ and $a_2(\omega_1)$, respectively. We only consider the case of $|a_{01}| = |a_{02}|$ and $|a_1| = |a_2|$ for simplicity. The balance requirement is usually satisfied in experiments[6-8].

The input-output Heisenberg evolutions of the field modes of the NOPA operating in the state of amplification (The injected subharmonic signals, $a_{01}(\omega_1)$ and $a_{02}(\omega_1)$, and harmonic pump field $a_0(\omega_0)$ are in phase.) are given by[13]

$$X_{a1} = X_{01} \cosh r + X_{02} \sinh r$$
$$Y_{a1} = Y_{01} \cosh r - Y_{02} \sinh r$$
$$X_{a2} = X_{02} \cosh r + X_{01} \sinh r \quad (1)$$
$$Y_{a2} = Y_{02} \cosh r - Y_{01} \sinh r,$$

where, $X_{a1}$, $X_{a2}$ ($Y_{a1}$, $Y_{a2}$) denote the amplitude quadrature components (phase quadrature components) of the two output modes and $X_{01}$, $X_{02}$ ($Y_{01}$, $Y_{02}$) are the corresponding quadrature components of the two injected fields. $r(0 \leq r < \infty)$ is the squeezing factor which depends on the length and the effective second-order susceptibility of the nonlinear crystal used for NOPA, the losses of the optical cavity, as well as the intensity of pump field. From Eqs.(1), we can easily calculate the variances of the difference of amplitude quadratures and the sum of phase quadratures between $a_1(\omega_1)$ and $a_2(\omega_1)$:

$$\langle \delta^2(X_{a1} - GX_{a2}) \rangle = \langle \delta^2(Y_{a1} + GY_{a2}) \rangle = \frac{1}{\cosh(2r)} \quad (2)$$

where the variances have been normalized to the shot noise limit(SNL) of the total beams $a_1(\omega_1)$ and $a_2(\omega_1)$, G is the optimum gain factor[13].

Then, one of the EPR beams ($a_1$) is injected in an other optical cavity involving a nonlinear $\chi^{(2)}$ crystal (SFG). $b_1^{in}(\omega_1) = a_1(\omega_1)$ and $b_2^{in}(\omega_2)$ are the injected signal and pump field ($\omega_1 \neq \omega_2$) of the SFG, respectively. $b_1(\omega_1)$ and $b_2(\omega_2)$ stand for the corresponding intracavity-fields of $b_1^{in}$ and $b_2^{in}$. An output field $b_3$ at the frequency $\omega_3 = \omega_1 + \omega_2$ is generated via a nonlinear sum-frequency process in SFG cavity. We will analyze the process of SFG and discuss the preserved entanglement between the output field $b_3^{out}(\omega_3)$ of SFG and the retained $a_2(\omega_1)$ in next sections.

## 3 Sum-frequency generation

We calculate the quantum fluctuation characteristics of the SFG in an optical cavity using the semi-classical approach. The dynamics of small field fluctuations is described by linearizing the classical equations of motion in the vicinity of the stationary state. We consider that these field fluctuations are driven by the fluctuations of the input fields (including vacuum field) through the coupling mirrors. It has been theoretically demonstrated in Ref [14] that the semi-classical approach can lead to the same results with the standard quantum methods [15] for the quantum fluctuations of the output fields from an optical cavity.

Under the case of perfect phase matching, zero detuning, small one-pass gain and small losses, the equations of motion for the classical amplitudes $\beta_1, \beta_2, \beta_3$ of intracavity fields associated with the annihilation operators $b_1, b_2, b_3$ can be expressed by[16]

$$\tau \dot{\beta}_1(t) = -(\gamma_1 + \rho_1)\beta_1(t) + \chi \beta_2^*(t)\beta_3(t) + \sqrt{2\gamma_1}\beta_1^{in}(t) + \sqrt{2\rho_1}c_1^{in}(t)$$
$$\tau \dot{\beta}_2(t) = -(\gamma_2 + \rho_2)\beta_2(t) + \chi \beta_1^*(t)\beta_3(t) + \sqrt{2\gamma_2}\beta_2^{in}(t) + \sqrt{2\rho_2}c_2^{in}(t) \quad (3)$$
$$\tau \dot{\beta}_3(t) = -(\gamma_3 + \rho_3)\beta_3(t) - \chi \beta_1(t)\beta_2(t) + \sqrt{2\gamma_3}\beta_3^{in}(t) + \sqrt{2\rho_3}c_3^{in}(t),$$

where, the round-trip time $\tau$ of light in the cavity is assumed to be same for all the three fields. The $\gamma_i$ and $\rho_i$ ($i=1,2,3$) stand for the single pass loss parameters corresponding to the transmission of the input and output couplers of the cavity and extra intracavity losses, respectively. The $\gamma_i$ are directly related to the amplitude refection and the transmission coefficients of the input and the output couplers of the optical cavity and the $\rho_i$ to the amplitude transmission coefficient of the optical medium in the cavity. In Eqs.(3), $\gamma_i$ and $\rho_i$ express the losses during the single pass in the cavity, and is not the losses in an unit time as usual, thus the round-trip time $\tau$ appears in the equations[16]. The $\beta_i^{in}$ ($i=1,2,3$) are the classical amplitudes of $b_i^{in}$ ($b_3^{in}$ is vacumm field). The $c_i^{in}$ ($i=1,2,3$) denote the extra noise amplitudes in addition to the intracavity field $b_i$ due to the internal loss mechanism. $\chi$ is the effective nonlinear coupling parameter, which is proportional to the second order susceptibility of the medium.

Assuming that the pump field $b_2$ is strong and can be considered to be undepleted, we have $\beta_2 = \beta_2^* = E$ and we can linearize the evolution equations around the mean amplitudes:

$$\begin{aligned} \beta_1(t) &= \langle \beta_1 \rangle + \delta\beta_1(t) & \beta_1^{in}(t) &= \langle \beta_1^{in} \rangle + \delta\beta_1^{in}(t) \\ \beta_3(t) &= \langle \beta_3 \rangle + \delta\beta_3(t) & \beta_3^{in}(t) &= \delta\beta_3^{in}(t). \end{aligned} \quad (4)$$

Substituting Eqs.(4) into Eqs.(3), we obtain the fluctuation dynamics equations,

$$\begin{aligned} \tau\delta\dot{\beta}_1(t) &= -(\gamma_1+\rho_1)\delta\beta_1(t) + \chi E\delta\beta_3^*(t) + \sqrt{2\gamma_1}\delta\beta_1^{in}(t) + \sqrt{2\rho_1}c_1^{in}(t) \\ \tau\delta\dot{\beta}_3(t) &= -(\gamma_3+\rho_3)\delta\beta_3(t) - \chi E\delta\beta_1^*(t) + \sqrt{2\gamma_3}\delta\beta_3^{in}(t) + \sqrt{2\rho_3}c_3^{in}(t). \end{aligned} \quad (5)$$

After Fourier transformation, we have

$$(i\omega\tau + \gamma_1 + \rho_1)\delta\beta_1(\omega) = \chi E \delta\beta_3(\omega) + \sqrt{2\gamma_1}\delta\beta_1^{in}(\omega) + \sqrt{2\rho_1}c_1^{in}(\omega)$$
$$(i\omega\tau + \gamma_3 + \rho_3)\delta\beta_3(\omega) = -\chi E \delta\beta_1(\omega) + \sqrt{2\gamma_3}\delta\beta_3^{in}(\omega) + \sqrt{2\rho_3}c_3^{in}(\omega),$$
(6)

where $\omega$ is the analysis frequency.

Using the boundary condition on the output coupling mirror[17]

$$\delta\beta_3^{out} = \sqrt{2\gamma_3}\delta\beta_3 - \delta\beta_3^{in},$$
(7)

we obtain the fluctuation of output field $b_3^{out}$ in term of the input fluctuation,

$$\delta\beta_3^{out}(\omega) = \frac{1}{(i\omega\tau + \gamma_3 + \rho_3)(i\omega\tau + \gamma_1 + \rho_1) + (\chi E)^2}$$
$$\{[(i\omega\tau + \gamma_1 + \rho_1)(-i\omega\tau + \gamma_3 - \rho_3) - (\chi E)^2]\delta\beta_3^{in}(\omega) +$$
$$2\sqrt{\gamma_3\rho_3}(i\omega\tau + \gamma_1 + \rho_1)c_3^{in}(\omega) - 2\chi E\sqrt{\gamma_1\gamma_3}\delta\beta_1^{in}(\omega) - 2\chi E\sqrt{\rho_1\gamma_3}c_1^{in}(\omega)]\}.$$
(8)

## 4 Entanglement characteristics

From Eq.(8) and the definitions of the amplitude and phase quadratures, $X = \frac{1}{2}(b + b^+)$ and $Y = \frac{1}{2i}(b - b^+)$, the fluctuation spectra of the quadrature components of $b_3^{out}$ are calculated:

$$\delta X_{b3}^{out} = \frac{1}{R}[A\delta X_{b1}^{in} + B\delta Y_{b1}^{in} + C\delta X_{b3}^{in} + D\delta Y_{b3}^{in} + GX_{c3}^{in} + HY_{c3}^{in} + MX_{c1}^{in} + NY_{c1}^{in}]$$
$$\delta Y_{b3}^{out} = \frac{1}{R}[A\delta Y_{b1}^{in} - B\delta X_{b1}^{in} + C\delta Y_{b3}^{in} - D\delta X_{b3}^{in} + GY_{c3}^{in} - HX_{c3}^{in} + M\delta Y_{c1}^{in} - N\delta X_{c1}^{in}],$$
(9)

where

$$R = [(\gamma_1 + \rho_1)(\gamma_3 + \rho_3) - (\omega\tau)^2 + (\chi E)^2]^2 + [\omega\tau(\gamma_1 + \rho_1 + \gamma_3 + \rho_3)]^2$$

$$A = -2\chi E\sqrt{\gamma_1\gamma_3}[(\gamma_1 + \rho_1)(\gamma_3 + \rho_3) - (\omega\tau)^2 + (\chi E)^2]$$

$$B = -2\chi E\sqrt{\gamma_1\gamma_3}\omega\tau(\gamma_1 + \rho_1 + \gamma_3 + \rho_3)$$

$$C = [(\gamma_1 + \rho_1)(\gamma_3 - \rho_3) + (\omega\tau)^2 - (\chi E)^2][(\gamma_1 + \rho_1)(\gamma_3 + \rho_3) - (\omega\tau)^2 + (\chi E)^2]$$
$$+ (\omega\tau)^2(\gamma_1 + \rho_1 + \gamma_3 + \rho_3)(\gamma_3 - \rho_3 - \gamma_1 - \rho_1)$$

$$D = \omega\tau[(\gamma_1 + \rho_1)(\gamma_3 - \rho_3) + (\omega\tau)^2 - (\chi E)^2](\gamma_1 + \rho_1 + \gamma_3 + \rho_3)$$
$$- \omega\tau[(\gamma_1 + \rho_1)(\gamma_3 + \rho_3) - (\omega\tau)^2 + (\chi E)^2](\gamma_3 - \rho_3 - \gamma_1 - \rho_1)$$

$$G = 2\sqrt{\gamma_3\rho_3}\{[(\gamma_1 + \rho_1)(\gamma_3 + \rho_3) + (\chi E)^2](\gamma_1 + \rho_1) + (\omega\tau)^2(\gamma_3 + \rho_3)\}$$

$$H = 2\sqrt{\gamma_3\rho_3}(\omega\tau)[(\gamma_1 + \rho_1)^2 + (\omega\tau)^2 - (\chi E)^2]$$

$$M = -2\chi E\sqrt{\rho_1\gamma_3}[(\gamma_1 + \rho_1)(\gamma_3 + \rho_3) - (\omega\tau)^2 + (\chi E)^2]$$

$$N = -2\chi E\sqrt{\rho_1\gamma_3}(\omega\tau)(\gamma_1 + \rho_1 + \gamma_3 + \rho_3)$$

$X_{bi}^{in}, Y_{bi}^{in}$ $(i = 1,3)$ and $X_{ci}^{in}, Y_{ci}^{in}$ $(i = 1,3)$ denote the amplitude and phase quadratures of $b_1^{in}, b_3^{in}$ and $c_1^{in}, c_3^{in}$, respectively.

For observing the optimum entanglement between $b_3^{out}$ and $a_2$, we should implement the appropriate unitary transformation on $a_2$, in which the quantum properties of the optical field $a_2$ will not be changed. Generally, the amplitude and phase quadratures of $a_2$ are expressed by[18]

$$X_{a2}^\theta = \frac{1}{2}(a_2 e^{-i\theta} + a_2^+ e^{i\theta}), Y_{a2}^\theta = \frac{1}{2i}(a_2 e^{-i\theta} - a_2^+ e^{i\theta}), \tag{10}$$

where, $\theta$ is the phase angle of $X_{a2}^\theta$ and $Y_{a2}^\theta$ rotated from the initial $X_{a2}$ and $Y_{a2}$, and it can be conveniently completed by adjusting the phase of the local oscillator or using a phase-shifter in experiments.

Rewriting Eqs.(10) in term of the amplitude and phase quadratures $X_{a2}$ and $Y_{a2}$ of $a_2$, we obtain

$$X_{a2}^{\theta} = X_{a2}\cos\theta + Y_{a2}\sin\theta, \tag{11}$$

and

$$Y_{a2}^{\theta} = -X_{a2}\sin\theta + Y_{a2}\cos\theta. \tag{12}$$

The correlation fluctuations of amplitude and phase quadratures between $b_3^{out}$ and $a_2^{\theta}$ ($a_2^{\theta}$ is the transformed $a_2$ according to Eqs.(10)) are expressed by:

$$\begin{aligned}\delta X_{b3}^{out} - g\delta X_{a2}^{\theta} &= \frac{1}{R}[A\delta X_{b1}^{in} + B\delta Y_{b1}^{in} + C\delta X_{b3}^{in} + D\delta Y_{b3}^{in} + G X_{c3}^{in} + H Y_{c3}^{in} + M X_{c1}^{in} + N Y_{c1}^{in}] \\ &\quad - g\delta X_{a2}\cos\theta - g\delta Y_{a2}\sin\theta \\ &= (\frac{A}{R}\delta X_{b1}^{in} - g\cos\theta\delta X_{a2}) + (\frac{B}{R}\delta Y_{b1}^{in} - g\sin\theta\delta Y_{a2}) \\ &\quad + \frac{1}{R}[C\delta X_{b3}^{in} + D\delta Y_{b3}^{in} + G X_{c3}^{in} + H Y_{c3}^{in} + M X_{c1}^{in} + N Y_{c1}^{in}],\end{aligned} \tag{13}$$

$$\begin{aligned}\delta Y_{b3}^{out} + g\delta Y_{a2}^{\theta} &= \frac{1}{R}[A\delta Y_{b1}^{in} - B\delta X_{b1}^{in} + C\delta Y_{b3}^{in} - D\delta X_{b3}^{in} - H X_{c3}^{in} + G Y_{c3}^{in} + M\delta Y_{c1}^{in} - N\delta X_{c1}^{in}] \\ &\quad - g\delta X_{a2}\sin\theta + g\delta Y_{a2}\cos\theta \\ &= (\frac{A}{R}\delta Y_{b1}^{in} + g\cos\theta\delta Y_{a2}) - (\frac{B}{R}\delta X_{a1}^{in} + g\sin\theta\delta X_{a2}) \\ &\quad + \frac{1}{R}[-D\delta X_{b3}^{in} + C\delta Y_{b3}^{in} - H X_{c3}^{in} + G Y_{c3}^{in} - N X_{c1}^{in} + M Y_{c1}^{in}],\end{aligned} \tag{14}$$

where, $g$ is an adjustable gain factor. Since $b_1^{in} = a_1$ and the quantum fluctuation is not changed in the unitary transformation, we may substitute Eqs.(1) into Eqs.(13) and (14), and get

$$\begin{aligned}\delta X_{b3}^{out} - g\delta X_{a2}^{\theta} &= [\frac{A}{R}\cosh r - g\cos\theta\sinh r]X_{01} + [\frac{A}{R}\sinh r - g\cos\theta\cosh r]X_{02} \\ &\quad + [\frac{B}{R}\cosh r + g\sin\theta\sinh r]Y_{01} - [\frac{B}{R}\sinh r + g\sin\theta\cosh r]Y_{02} \\ &\quad + \frac{1}{R}[C\delta X_{b3}^{in} + D\delta Y_{b3}^{in} + G X_{c3}^{in} + H Y_{c3}^{in} + M X_{c1}^{in} + N Y_{c1}^{in}],\end{aligned} \tag{15}$$

and

$$\delta Y_3^{out} + g\delta Y_2^{\theta} = [\frac{A}{R}\cosh r - g\cos\theta \sinh r]Y_{01} - [\frac{A}{R}\sinh r - g\cos\theta \cosh r]Y_{02}$$
$$-[\frac{B}{R}\cosh r + g\sin\theta \sinh r]X_{01} - [\frac{B}{R}\sinh r + g\sin\theta \cosh r]X_{02} \quad (16)$$
$$+\frac{1}{R}[-D\delta X_{b3}^{in} + C\delta Y_{b3}^{in} - HGX_{c3}^{in} + GY_{c3}^{in} - NX_{c1}^{in} + MY_{c1}^{in}].$$

Then, the correlation variance of the difference of amplitude quadratures and the sum of phase quadratures are obtained

$$\langle \delta^2(X_{b3}^{out} - gX_{a2}^{\theta})\rangle = \langle \delta^2(Y_{b3}^{out} + gY_{a2}^{\theta})\rangle$$
$$= [\frac{A}{R}\cosh r - g\cos\theta \sinh r]^2 + [\frac{A}{R}\sinh r - g\cos\theta \cosh r]^2$$
$$+ [\frac{B}{R}\cosh r + g\sin\theta \sinh r]^2 + [\frac{B}{R}\sinh r + g\sin\theta \cosh r]^2 \quad (17)$$
$$+ \frac{1}{R^2}[C^2 + D^2 + G^2 + H^2 + M^2 + N^2]$$
$$= \frac{4(\chi E)^2 \gamma_1 \gamma_3}{R}\frac{e^{2r} + e^{-2r}}{2} - \frac{A}{R}g\cos\theta(e^{2r} - e^{-2r})$$
$$+ g^2 \frac{e^{2r} + e^{-2r}}{2} + \frac{B}{R}g\sin\theta(e^{2r} - e^{-2r})$$
$$+ \frac{1}{R^2}[C^2 + D^2 + G^2 + H^2 + M^2 + N^2].$$

Taking $\frac{A}{\sqrt{R}} = 2(\chi E)\sqrt{\gamma_1 \gamma_3}\cos\varphi$, and $\frac{B}{\sqrt{R}} = 2(\chi E)\sqrt{\gamma_1 \gamma_3}\sin\varphi$, we have

$$\langle \delta^2(X_{b3}^{out} - gX_{a2}^{\theta})\rangle = \langle \delta^2(Y_{b3}^{out} + gY_{a2}^{\theta})\rangle =$$
$$\frac{4(\chi E)^2 \gamma_1 \gamma_3}{R}\frac{e^{2r} + e^{-2r}}{2} - (2\chi E\sqrt{\gamma_1 \gamma_3})\frac{1}{\sqrt{R}}g\cos(\varphi + \theta)(e^{2r} - e^{-2r}) \quad (18)$$
$$+ g^2 \frac{e^{2r} + e^{-2r}}{2} + \frac{1}{R^2}[C^2 + D^2 + G^2 + H^2 + M^2 + N^2].$$

Calculating the minimum value of Eq.(18) in term of $g$, we obtain the optimum gain:

$$g_{opt} = \frac{2(\chi E)\sqrt{\gamma_1 \gamma_3}}{\sqrt{R}} \cos(\varphi + \theta) \frac{e^{2r} - e^{-2r}}{e^{2r} + e^{-2r}},$$

and the corresponding correlation variance equals to

$$S = \left\langle \delta^2(X_{b3}^{out} - g_{opt} X_{a2}^{\theta}) \right\rangle_{min} = \left\langle \delta^2(Y_{b3}^{out} + g_{opt} Y_{a2}^{\theta}) \right\rangle_{min} = \tag{19}$$

$$\frac{2(\chi E)^2 \gamma_1 \gamma_3}{R} [e^{2r} + e^{-2r} - \cos^2(\theta + \varphi) \frac{(e^{2r} - e^{-2r})^2}{e^{2r} + e^{-2r}}] + \frac{1}{R^2}[C^2 + D^2 + G^2 + H^2 + M^2 + N^2],$$

when $\theta = -\varphi$ $S$ reaches the minimum value $S_{min}$

Fig.2 shows the minimum correlation fluctuation spectra $S_{min}$ versus the normalized analysis frequency ($\Omega = \omega \tau / \gamma_1$) for different initial squeezing factor $r$. Obviously, at zero frequency ($\Omega = 0$), the maximum correlation is obtained ($S_{min}(\Omega = 0)$ reaches the minimum), and the larger the initial squeezing factor $r$ is, the better the preserved correlation between $a_2$ and $b_3^{out}$ is

In Fig.3 and Fig.4, the dependences of $S_{min}$ on the pump parameters ($\chi E / \gamma_1$) are calculated for different relative transmissions of $\gamma_3 / \gamma_1$ and different squeezing factors $r$, respectively. For the smaller pump parameters, the smaller $\gamma_3 / \gamma_1$ is better ($S_{min}$ is smaller), but for larger pump parameters the larger $\gamma_3 / \gamma_1$ value corresponds to smaller $S_{min}$. For a given $\gamma_3 / \gamma_1$, we have an optimal pump parameter at which the $S_{min}$ reaches to a minimum. If $\gamma_3 / \gamma_1$ increases, the optimal pump parameter increases too. That is because for larger output transmission $\gamma_3$ of $b_3^{out}$, the higher pump power is needed to achieve the optimal SFG. Fig.4 shows, when the parameters of SFG cavity and initial squeezing are given, we should choose the optimal pump power to meet the smallest $S_{\min}$ value for successfully preserving the quantum correlation. In the ideal limit without any intracavity losses, if taking $g_{opt} = 1$, the $S_{\min}$ will equal to the initial EPR correlation between $a_1$ and $a_2$. In this case $\theta = \pi$, it means that there is a phase difference of $\pi$ between the input ($X_{a1} = X_{b1}^{in}$) and the output ($X_{b3}^{out}$) of SFG[19], thus when we measure the correlation between $X_{b3}^{out}$ and $X_{a2}$, a phase shift of $\pi$ should be added on $X_{a2}$.

In Fig.5, the relation of the minimum correlation fluctuations $S_{\min}$ versus the initial squeezing factor $r$ is drawn. The correlation variance $S_{\min}$ between the field $a_2$ and $b_3^{out}$ decrease, i.e. the entanlement increases when the $r$ increases. From Eq.(2) we know that the initial correlation variances of both amplitude and phase quadratures between $a_1$ and $a_2$ are smaller than the normalizied SNL for $r > 0$, thus the inseparability criterion of EPR entanglement state for continuous variables proposed by Duan[20] is satisfied, that is

$$\langle \delta^2(X_{a1} - GX_{a2}) \rangle + \langle \delta^2(Y_{a1} + GY_{a2}) \rangle < 2. \qquad (20)$$

In the case of $r > 0$, the minimum correlation variances for the amplitude and phase quadratures between $a_2$ and $b_3^{out}$ fields are equal (see Eq.(19)) and both smaller than 1 also(Fig.5), so we have

$$\langle \delta^2(X_{b3}^{out} - g_{opt} X_{a2}^{\theta}) \rangle_{\min} + \langle \delta^2(Y_{b3}^{out} + g_{opt} Y_{a2}^{\theta}) \rangle_{\min} < 2 \ .$$

It means, the correlation variables of the quadratures between the field $a_2$ and $b_3^{out}$ satisfy the inseparability criterion for quantum entangled state. Once the initial entanglement between the field $a_1$ and $a_2$ exists ($r>0$), the entanglement between the field $a_2$ and $b_3^{out}$ also exists. The better the initial entanglement is, the larger the remaining entanlement after the frequency conversion is. For an ideal SFG without the intracavity losses and taking $\Omega=0$ and $(\chi E)^2=\gamma_1\gamma_3$, the remaining correlation variances will equal to the initial variances from Eq.(19), thus the entanglement can be perfectly preserved. However, for any experimental system with the losses, the remaining entanglement is always worse than that of the initial state.

## 4 Conclusion

Our analyses theoretically proved that the initial EPR entanglement between the amplitude and phase quadratures of entangled beams can be preserved after the frequency of one of the beams is conversed via an intracavity nonlinear interaction of SFG. We calculated the dependences of the resultant correlation fluctuation spectra on the parameters of SFG system, the pump power and the initial squeezing. The squeezing parameter $r=0.6$ corresponds to the correlation fluctuation of the initial EPR beams is $\approx$ 5.2 dB below the SNL, which has been realized by experiments [6-8], in this case the preserved entanglement is about 2 dB below the SNL ($S_{\min}\approx 0.63$) in the experimentally accessible systems. The frequency conversion of entangled optical beams is important in constructing complete quantum communication networks. The calculated results may be a useful reference for the design of quantum communication systems.

Tan Aihong thanks J. Zhang for the helpful discussion. This research was supported in part by the National Natural Science Foundation of China(Grant No.60238010, 60378014), the Major State Basic Research Project of China(Grant No.2001CB309304).

*Email: changde@sxu.edu.cn

Fig.1 Scheme of physical system for QFC

**Fig.2 The correlation fluctuation spectra $S_{min}$ versus the normalized analysis frequency $\Omega$** $\Omega = \omega\tau/\gamma_1$

    Transmission of the output coupler $\gamma_3/\gamma_1 = 1$

    extra intracavity losses $\rho_1/\gamma_1 = \rho_3/\gamma_1 = 0.1$

  Pump parameter $\chi E/\gamma_1 = 1$ gain factor $g$ is chosen to be the optimum value

**Fig.3 The correlation fluctuation spectra $S_{min}$ at $\Omega = 0$ versus the Pump parameter $\chi E/\gamma_1$ for different relative transmission $\gamma_3/\gamma_1 = 0.6, 1, 1.4$**

    extra intracavity losses $\rho_1/\gamma_1 = \rho_3/\gamma_1 = 0.1$

    gain factor g is chosen to be the optimum value

    the squeezing parameter $r = 2$

**Fig.4 The correlation fluctuation spectra $S_{min}$ at $\Omega = 0$ versus the Pump parameter $\chi E/\gamma_1$ for different squeezing factor( $r = 0.6, 1, 2$ )**

    the normalized frequency $\Omega = 0$

    extra intracavity losses $\rho_1/\gamma_1 = \rho_3/\gamma_1 = 0.1$

    gain factor $g$ is chosen to be the optimum value

    Transmission of input-output couple $\gamma_3/\gamma_1 = 1$

**Fig.5 The minimum correlation fluctuation $S_{min}$ at $\Omega = 0$ versus the initial squeezing factor $r$**

    Transmission of the output coupler $\gamma_3/\gamma_1 = 1$

    extra intracavity losses $\rho_1/\gamma_1 = \rho_3/\gamma_1 = 0.1$

  Pump parameter $\chi E/\gamma_1 = 1$ gain factor $g$ is chosen to be the optimum value

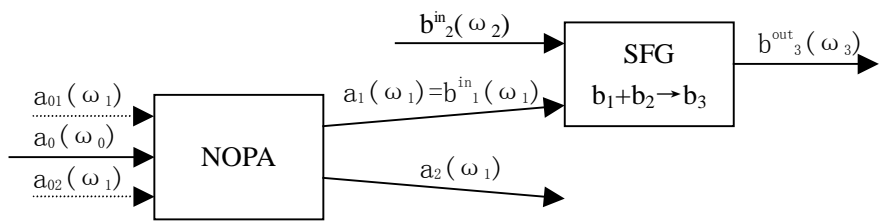

Fig. 1

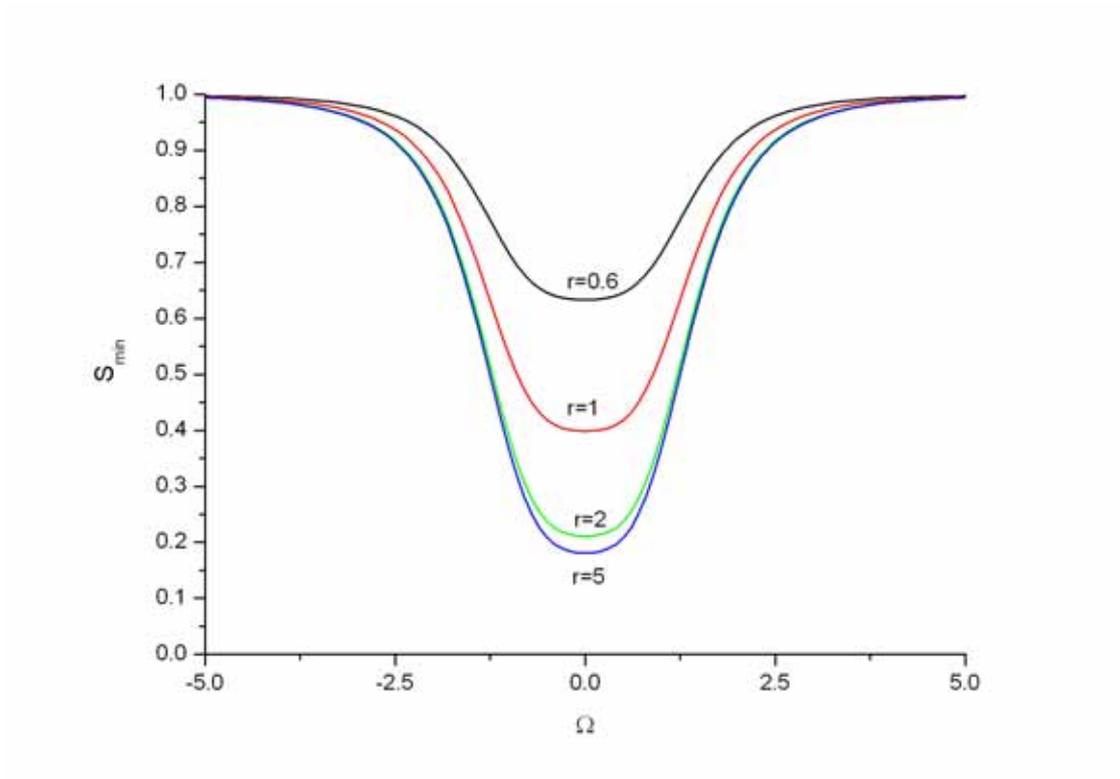

Fig.2

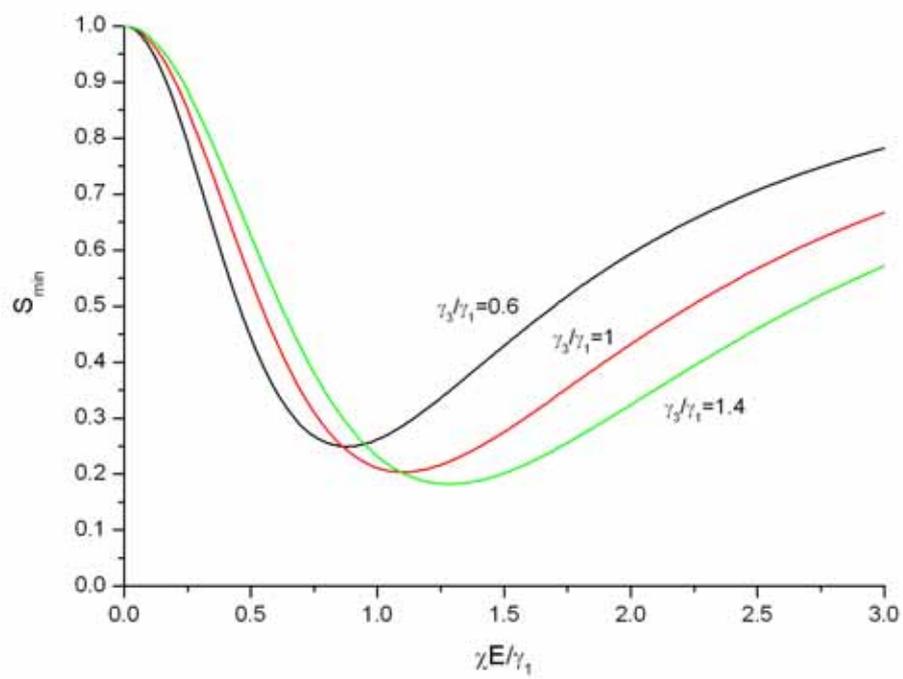

Fig. 3

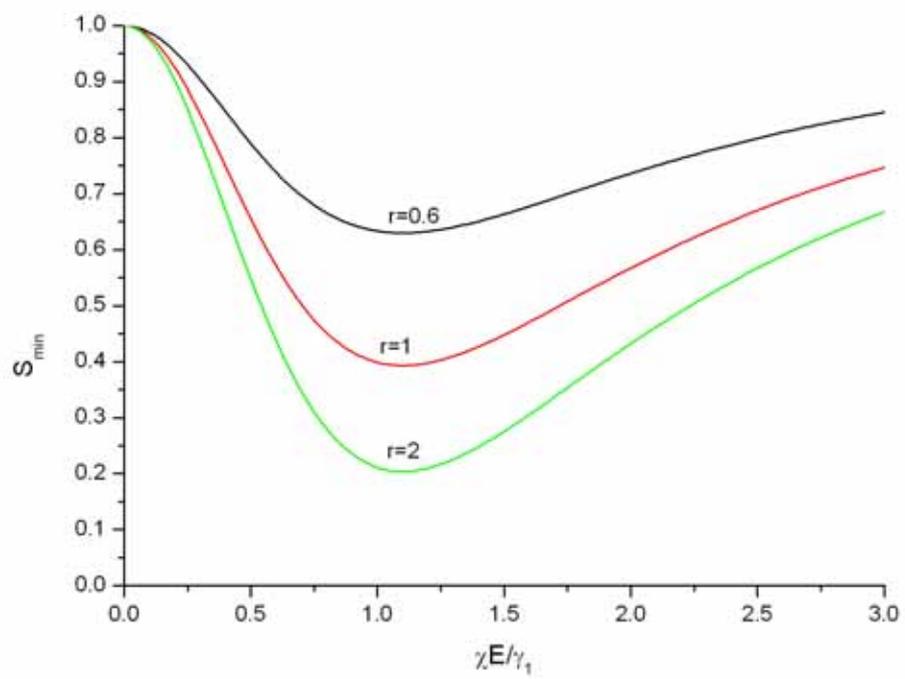

Fig. 4

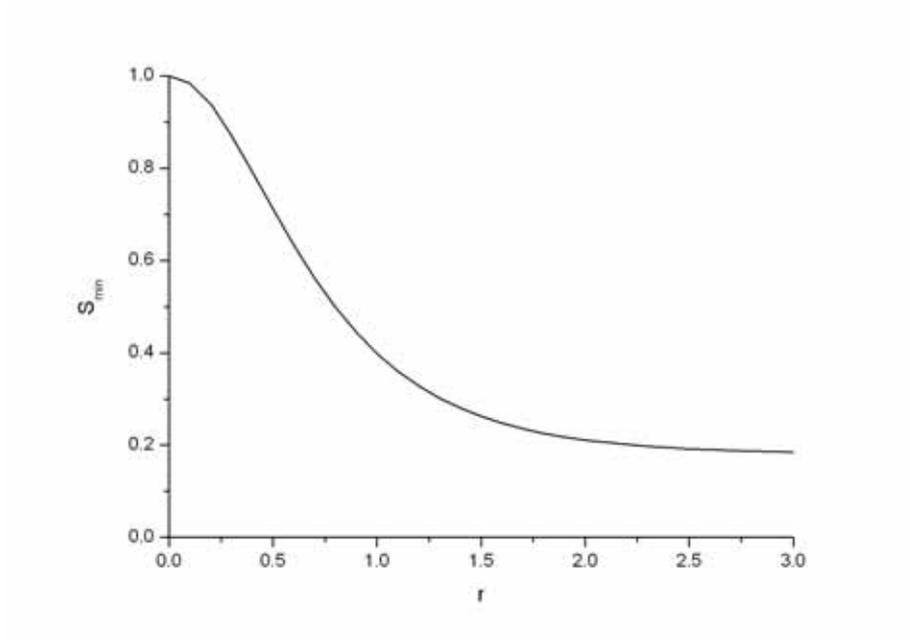

Fig. 5